\documentclass[letterpaper, bothsides, 12pt]{article} 
\usepackage[letterpaper,left=1.5cm,top=1.5cm,right=1.5cm,bottom=1.5cm]{geometry}  
\usepackage[spanish,english,activeacute]{babel}
\usepackage[utf8]{inputenc}
\usepackage{latexsym} 
\usepackage{layout}
\usepackage{float}
\usepackage{textcomp}
\usepackage{amstext}
\usepackage[pdftex]{graphicx}
\usepackage{multicol} 
\usepackage{longtable} 
\usepackage{lscape} 
\usepackage{multirow} 
\usepackage{array} 
\usepackage{amssymb}
\usepackage{amsmath} 
\usepackage{amsthm}
\usepackage[T1]{fontenc} 
\usepackage{listings} 
\usepackage[center, small]{caption2} 
\usepackage{graphicx}

\usepackage{titlesec}
\theoremstyle{definition}
\usepackage[pdftex]{color}
\usepackage[pdftex,colorlinks=true,linkcolor=blue,citecolor=black,urlcolor=cyan]{hyperref}

\renewcommand{\contentsname}{Contenido}
\renewcommand{\partname}{Parte}
\renewcommand{\indexname}{Lista Alfabética}
\renewcommand{\appendixname}{Apéndice}
\renewcommand{\figurename}{Figura}
\renewcommand{\listfigurename}{Lista de Figuras}
\renewcommand{\listtablename}{Lista de Tablas}
\renewcommand{\abstractname}{Resumen}

\newlength{\LL}

\title{ \bf Different electromagnetic physical representations of the Dirac's oscillator according with its spatial dimension}
\author{J.S.Montañez-Moyano \footnote{jsmontanezm@unal.edu.co}, C.J. Quimbay Herrera \footnote{cjquimbayh@unal.edu.co}  \\
Departamento de F\'{i}sica, Universidad Nacional de Colombia\\ Sede Bogotá.}
\date{\today}

\begin{document}

\renewcommand{\contentsname}{Contenido}
\renewcommand{\partname}{Parte}
\renewcommand{\indexname}{Lista Alfabética}
\renewcommand{\appendixname}{Apéndice}
\renewcommand{\figurename}{Figura}
\renewcommand{\listfigurename}{Lista de Figuras}
\renewcommand{\tablename}{Tabla}
\renewcommand{\listtablename}{Lista de Tablas}
\renewcommand{\abstractname}{Abstract}
\renewcommand{\refname}{References}

\maketitle

\begin{abstract}

\noindent Dirac's oscillator (DO) is one of the most studied systems in the Relativistic Quantum Mechanics and in the physical-mathematics. In particular, we show that this system has an unique property which it has not ever seen in other known systems: According to its spatial dimensionality, DO represent  physical systems with very different electromagnetic nature. So far in the literature, it has been proved using the covariant method the gauge invariance of the Dirac's oscillator potential. It has also shown that in (3+1)dimensions the DO represents a relativistic and electrically neutral fermion with magnetic dipole momentum, into a dielectric medium with spherical symmetry and under the effect of an electric field which depends of the radial distance. In this work,and using the same methodology, we show that (2+1) dimensional DO represents a 1/2-spin relativistic fermion under the effect of a uniform and perpendicular external magnetic field; whereas in (1+1) dimensions DO reproduces a relativistic and electrically charged fermion interacting with a linear electric field. Additionally, we prove that DO does not have chiral invariance, independent of its dimensionality, due to the interaction potential which breaks explicitly the chiral symmetry $U(1)_R \times U(1)_L$ but it preserves the global gauge symmetry $U(1)$.      
\end{abstract}

\section{Introduction}

\noindent Dirac's oscillator (DO) was introduced by \cite{1} as relativistic analogy of the Quantum Harmonic Oscillator. The term related to the potential is a linear expression in its position, unlike that the non relativistic case where this potential depends as $r^2$, because Dirac's equation is a linear expression both in its spatial derivatives and temporal one (in the Schrödinger's equation, temporal derivative is linear but it has a Laplacian instead a gradient). 

\vspace{0.5cm}

\noindent This physical system has been widely studied since it allows to study several aspects both the theoretical and experimental physics. For instance in \cite{2} it is used as an element to show that DO contains elements which give a value to the Supersymmetry Quantum Mechanics because, in other aspects, the solution of the system's associated equation does not have a Supersymmetry's breaking due to the topology of the potential and the Supersymmetric Hamiltonian can be constructed in order to study the physical system. Also, in the same reference, it is said that DO has applications in the QCD because they can build Quarks' confinement models since the interaction potential in both cases have a linear potential.

\vspace{0.5cm }

\noindent Also, the relation between the DO in (2+1) dimensions and the monolayer graphene has been studied in \cite{2} \cite{4} \cite{5} through the left-chirality due to the Spin-Orbit coupling which can be understood as an internal magnetic field and it has a direct relation with the linear term of the DO since $m\omega=eB_I$ \cite{4}. Dirac's oscillator has been studied from the Jaynes-Cumming model \cite{6}, non-relativistic limit \cite{7}, its exact solutions in polar coordinates in (2+1) dimensions \cite{8}, from the Relativistic Quantum Mechanics \cite{9}, the canonical quantization in (1+1) and (3+1) dimensions \cite{10}, the magnetic condensate for the flat Dirac's oscillator \cite{11}.

\vspace{0.5cm}

\noindent By the other hand, several physical systems has been studied in multiples dimensions (1+1), (2+1) and (3+1) \footnote{With one spatial dimension and one temporal; with 2 spatial dimensions; and three spatial dimensions and one temporal, respectively}. It has been seen that, for instance, superfluid behaviour changes between (1+1) and (2+1) dimensions  because in the first case there are not phase quantum fluctuations which allows the generation of Bose-Einstein condensate; whereas in (2+1) dimensions it has been observed these variations that allow that the condensate production is being possible \cite{19}. In \cite{20} is being studied the behaviour of (2+1) dimensional gas which behaves as a bosonic gas at zero temperature in an anisotropic optical net, it makes a transition to a (1+1) dimensional system through a external potential that has the form $\sum_{\alpha=x,y,z} V_{\alpha} \cos ^2 \left(k\alpha \right)$. After it made this dimensional reduction, the physical system behaves as Mott's insulator after it reached a critical value. 

\vspace{0.5cm}

\noindent In the case of \cite{21} we initially see a phase transition from an insulator superfluid with magnons(SIT)into a quantum ferromagnetic with spin interactions and then it behaves as superposition of singlet and triplet states which obeys a powers law through the transversal direction of the spontaneous magnetization. On the other hand, in \cite{22} is studied optical solitons in (1+1) and (2+1) dimensions, they obtain that the difference between both systems is in their integrability, since in (1+1) case it can be solved using Fan expressions which contains solutions with sinusoidal and cosenoidal terms; whereas in (2+1) system, it requires to use a semivariational principle and numerical methods. In the case of \cite{23} and \cite{24}, they solve Schrödinger equation in (2+1) and (3+1) dimensions in systems with strong correlations in bosons and their relation with Mott's isolator. 

\vspace{0.5cm}

\noindent In this way, we observe that in systems such as they are described by \cite{9} \cite{19} \cite{20} \cite{21} \cite{22} \cite{23} \cite{24} \cite{30}, there are different behaviours in the physical system when their dimensionalities change.  Dirac's oscillator in (1+1), (2+1) and (3+1) dimensions is a system that changes when its dimensions are modified \footnote{In this case, it corresponds to the Dirac's equation in presence of a linear interaction term which is well known as  Dirac's oscillator}. At the same time, physical interpretation of Dirac's oscillator in (3+1) dimensions has been made by \cite{9} as a particle without charge with an anomalous magnetic moment interacting with a linear static electric field. They also calculate their energy spectra (eigenvalues) and wavefunctions (eigenfunctions). One of their conclusions in this work is that spin-orbit coupling term only contributes as a constant in the energy spectra of this system (in other words, its spectra is independent of the radial coordinate  $r$) and, also, in its non relativistic limit and for this coordinate we obtain the non-relativistic harmonic oscillator in (3+1) dimensions.

\vspace{0.5cm}

\noindent On the other hand, chirality is a property that generally means: the structural characteristic of a finite system (molecule, atom or particle) that makes it impossible to superimpose it on its mirror image \cite{55}. Chiral symmetry has been studied by \cite{31} in order to do nonlinear realizations of chiral symmetry, chiral symmetry breaking in \cite{32} \cite{33} \cite{34}, spontaneous chiral symmetry breaking \cite{35} \cite{36} \cite{37} and dynamical chiral symmetry breaking \cite{38} \cite{39} \cite{40}. In this work, we shall demonstrate that (2+1) and (1+1) dimensional Dirac's oscillator preserves local gauge symmetry $U(1)$ but not local chiral symmetry $U(1)_R \times U(1)_L$ for a massless fermion.   

\vspace{0.5cm}

\noindent The main goal of this paper is to interpret physically DO in (2+1) and (1+1) dimensions, relating this system with well known electrodynamics systems and then we can notice that DO represents different physical systems when its dimensionality is changed. In this way, this work have the following structure: 1. Physical interpretation of (2+1) dimensional DO, 2. Physical interpretation of (1+1) dimensional DO, 3. Lagrangian density of the QED+DO in (2+1) dimensions in its symmetries $U(1)$ y $U(1)_R \times U(1)_L$ and, 4.Conclusions about these results.

\section{Physical interpretation of (2+1) dimensional Dirac's oscillator}

\noindent In order to know the physical representation of the Dirac's oscillator in (2+1) dimensions, let's suppose a particle with mass $m$ which its moving in a circular orbit with radius $R$ and with tangential velocity $v$. We know that its centripetal acceleration is given by $\frac{m v^2}{R}$ \cite{13} and it always is directed to the center of its orbit. On the other hand, it is also known that the force which is exerts by a magnetic field over a charged particle is always perpendicular to the particle's instantaneous velocity. Taking into account the previous information, now we consider the preceding particle which posses a electric charge $q$ and it is moving in a perpendicular plane to a uniform magnetic field $B$. The magnetic force which acts over the particle has the magnitude  $f= q \ v \ B$ towards the center of its orbit, i.e.

\begin{equation}
qvB= \frac{m v^2}{R}.
\label{eqBfield1}
\end{equation}

\vspace{0.5cm}

\noindent Thus we can say that the magnetic field has a magnitude given by

\begin{equation}
B= \frac{m v}{q R}.
\label{eqBfield2}
\end{equation}

\noindent Taking into account that the angular frequency of a particle with circular uniform movement can be written as $\omega= v/R $ then we can rewrite the expression (\ref{eqBfield2}) for the magnetic field as follows

\begin{equation}
B= \frac{m \omega}{q}.
\label{eqBfield3}
\end{equation}

\noindent Using the relation (\ref{eqBfield3}) is possible to verify that the magnetic field which acts over the particle is uniform since all the involved variables on it are parameters in this situation. Now, we consider a uniform conducting plate interacting with a perpendicular magnetic field to it, i.e. \cite{2} \cite{13} \cite{12}  

\begin{equation}
\vec{B}= - B \hat{z},
\label{eqE40}
\end{equation}

\noindent whereas the electric field in all the space is null, in other words

\begin{equation}
\vec{E}=0.
\label{eqE41}
\end{equation}

\noindent Now we suppose that the plane is big enough to neglect the edge effects. We will show that a Dirac's fermion interacting with the electromagnetic field produced by the plane lead us to the hamiltonian of the (2+1) dimensional Dirac's oscillator \cite{4}. In the same way that in (3+1) dimensional case, we shall show the Lorentz's covariant properties of the potential which describes this situation. 

\vspace{0.5cm}

\noindent We study the fermion from the reference frame proper of the plate, which we call laboratory reference system. We can calculate the electric and magnetic fields using the associated 4-electromagnetic potential, $A_{\mu}$. A expression that could describe this situation is \cite{11} 

\begin{equation}
A^{\mu}= \rho \left(0, y, -x \right),
\label{eqE42}
\end{equation}

\noindent where $\rho$ is a constant. Remembering that the classic expressions for the scalar and vectorial potential are given by \cite{13}

\begin{equation}
\vec{E}= - \vec{\nabla} \phi -\partial_t \vec{A},
\label{eqE4}
\end{equation}

\begin{equation}
\vec{B}= \vec{\nabla}\times \vec{A}.
\label{eqE5}
\end{equation}

\noindent Now, we shall to verify that 4-electromagnetic potential(\ref{eqE42}) lead us to the electric and magnetic fields (\ref{eqE40}) and (\ref{eqE41}). For this purpose, we use the definitions(\ref{eqE4}) and (\ref{eqE5}) and starting off with (\ref{eqE42}), we have

\begin{equation}
\vec{E}= - \vec{\nabla} \left(0 \right) - \rho \partial_t \left(y ,-x \right) \Longrightarrow \vec{E}= 0.
\label{eqE43}
\end{equation}

\begin{equation}
\vec{B}= \vec{\nabla} \times \vec{A} \Longrightarrow \vec{B}= \rho \left| \begin{array}{ccc}
\hat{x}& \hat{y}& \hat{z}\\
\partial_{x}& \partial_{y}&\partial_{z}\\
y& -x&0\\
\end{array}\right| \Longrightarrow \vec{B}= -2\rho \hat{z} \hspace{1 cm}\therefore \hspace{0.5cm} 2 \rho= B .
\label{eqE44}
\end{equation}

\noindent We observe that the expressions (\ref{eqE43}) and (\ref{eqE44}) are the same that (\ref{eqE40}) and (\ref{eqE41}), and for that reason we can conclude that the 4-electromagnetic potential given by (\ref{eqE42}) gives us the same physical conditions of the physical situation studied and previously defined. Now, the expression(\ref{eqE43}) have to have a Lorentz covariant form, so we use the gauge invariance of the electromagnetic fields to rewrite the 4-electromagnetic potential (\ref{eqE42}) in terms of the gauge function which have the following form

\begin{equation}
\Lambda= - \frac{\rho}{4} t x^2 - \frac{\rho}{4} t y^2 - \frac{\rho}{12} t^3 .
\label{eqE45}
\end{equation}

\noindent Now we rewrite the potential in terms of the gauge function $\Lambda$ as follows \cite{9}

\begin{equation}\nonumber
A_{\text{lab}}^{\mu \prime}= A_{\text{lab}}^{\mu} - \partial_{\mu} \Lambda
\end{equation}

\noindent So, first at all we obtain the temporal component, replacing (\ref{eqE42}) and (\ref{eqE45}) in the latter expression making $\mu=0$, i.e., 

\begin{eqnarray}\nonumber
A_{\text{lab}}^{0 \prime}&=& A_{\text{lab}}^{0} - \partial_{0} \Lambda , \\ \nonumber
&=& 0 - \partial_t \left[- \frac{\rho}{4} t x^2 - \frac{\rho}{4} t y^2 - \frac{1}{12} \rho t ^3 \right], \\ \nonumber
\end{eqnarray}

\begin{equation}
A_{\text{lab}}^{0 \prime}= \frac{\rho}{4} \left(x^2 + y^2 +t^2 \right).
\label{eqE46}
\end{equation}

\noindent Whereas that the spatial components are given by 

\begin{eqnarray}\nonumber
A_{\text{lab}}^{1 \prime}&=& A_{\text{lab}}^{0} - \partial_{1} \Lambda , \\ \nonumber
&=& \rho y - \partial_x \left( - \frac{\rho}{4} t x^2 - \frac{\rho}{4} t y^2 - \frac{1}{12} \rho t ^3 \right), \\ \nonumber
\end{eqnarray}

\begin{equation}
A_{\text{lab}}^{1 \prime}= \rho y + \frac{1}{2} \rho t x.
\label{eqE47}
\end{equation}

\begin{eqnarray}\nonumber
A_{\text{lab}}^{2 \prime}&=& A_{\text{lab}}^{2} - \partial_{2} \Lambda , \\ \nonumber
&=& - \rho x - \partial_y \left( - \frac{\rho}{4} t x^2 - \frac{\rho}{4} t y^2 - \frac{1}{12} \rho t ^3 \right), \\ \nonumber
\end{eqnarray}

\begin{equation}
A_{\text{lab}}^{2 \prime}= - \rho x + \frac{1}{2} \rho t y.
\label{eqE48}
\end{equation}

\noindent Taking into account  the expressions (\ref{eqE46}), (\ref{eqE47}) and (\ref{eqE48}) then the new 4-electromagnetic potential is 

\begin{equation}
A_{\text{lab}}^{\mu \prime}= \rho \left[- \frac{1}{4} \left(x^2 + y^2 + t^2 \right), y + \frac{1}{2} t x, -x + \frac{1}{2} t y\right].
\label{eqE49}
\end{equation}

\noindent Now, we check that the 4-potential (\ref{eqE49}) lead us to the same physical conditions for the electromagnetic field that we defined in (\ref{eqE40}) and (\ref{eqE41}). For that reason, we calculate the electric and magnetic field produced by (\ref{eqE49})

\begin{eqnarray}\nonumber
\vec{E}^{\prime}&=& - \vec{\nabla} \phi^{\prime} - \partial_t \vec{A}^{\prime} \\ \nonumber
&=& \frac{ \rho}{4} \vec{\nabla} \left(x^2 + y^2 + t^2 \right) - \rho \partial_t \left( y + \frac{1}{2} t x, -x + \frac{1}{2} t y \right)\\ \nonumber
&=& \frac{\rho}{4} \left(2x, 2y \right)- \rho \left(\frac{x}{2}, \frac{y}{2} \right),\\ \nonumber
\end{eqnarray}

\begin{equation}
\vec{E}^{\prime}= 0
\label{eqE50}
\end{equation}

\begin{eqnarray}\nonumber
\vec{B}^{\prime}&=& \vec{\nabla} \times \vec{A}^{\prime}\\ \nonumber
&=& \rho \left| \begin{array}{ccc}
\hat{x}& \hat{y}& \hat{z}\\
\partial_{x}& \partial_{y}&\partial_{z}\\
y + \frac{1}{2} t x& -x + \frac{1}{2} t y &0\\
\end{array}\right| \\ \nonumber
&=&\rho \left(-1 -1 \right)\hat{z}\\ \nonumber
&=&-2\rho \hat{z}
\end{eqnarray}

\begin{equation}
\vec{B}^{\prime}= -B \hat{z}
\label{eqE51}
\end{equation}

\noindent We can see that the expressions (\ref{eqE50}) and (\ref{eqE51}) are the same that (\ref{eqE40}) and (\ref{eqE41}), and for that reason we can conclude that the 4-electromagnetic potential (\ref{eqE49}) represents the same physical situation that we described initially. Now, we shall to do the coordinates transformation from the laboratory frame to fermion's frame by the introduction of the 4-velocity, $U^{\mu}= \left(1,\vec{0} \right)$, and for this purpose we use the definition \cite{9}

\begin{equation}
A^{\mu \prime}= \frac{\rho}{4} \left[2 \left(U \cdot x \right) x^{\mu} - x^2 U^{\mu} \right].
\label{eqAmu}
\end{equation}

\noindent Using the definitions 

\begin{equation}
U \cdot x= \left(1,0,0,0 \right) \cdot \left(t,r,0,0 \right) \Longrightarrow U \cdot x= t
\label{eqE21},
\end{equation}

\begin{equation}
x^2= x^{\mu} x_{\mu}= \left(t,r,0,0 \right) \cdot \left(t,-r,0,0 \right) \Longrightarrow x^2= t^2-r^2,
\label{eqE22}
\end{equation}

\noindent and then the 4-electromagnetic potential can be rewritten as follows

\begin{equation}
A^{\mu \prime}= \frac{\rho}{4} \left[2t x^{\mu} -\left(t^2 - x^2 - y^2 \right)U^{\mu} \right].
\label{eqE52}
\end{equation}

\noindent Whereas that the components of this 4-potential are

\begin{equation}
A^{0 \prime}= \frac{\rho}{4} \left(x^2 + y^2 + t^2 \right); \hspace{2cm} A^{1 \prime}= \frac{1}{2} \rho x t; \hspace{2cm} A^{2 \prime}= \frac{1}{2} \rho t y.
\label{eqE53}
\end{equation}

\noindent The results given by (\ref{eqE53}) are the same that the expressions(\ref{eqE46}),(\ref{eqE47}) and (\ref{eqE48}). By the other hand, being the 4-electromagnetic potential an explicit Lorentz covariant expression then the electromagnetic field tensor produced by the plate can be written as \cite{9}  

\begin{equation}
F_{\mu \nu}= \rho \left(U^{\mu} x^{\nu} - U^{\nu} x^{\mu} \right).
\label{FO2}
\end{equation}

\noindent The non-null components of this tensor are

\begin{equation}
F_{01}= \rho x; \hspace{1cm} F_{10}=- \rho x; \hspace{1cm} F_{02}= \rho y; \hspace{1 cm} F_{20}= -\rho y.
\label{eqE54}
\end{equation}

\noindent Now we shall study the Dirac fermion behaviour moving through the plate. For this goal, we take the interaction term \cite{15}

\begin{eqnarray}
\frac{1}{2} m \omega \sigma^{\mu \nu} F_{\mu \nu} = \frac{eB}{2} \sigma^{\mu \nu} F_{\mu \nu}.
\label{eqE54.1}
\end{eqnarray}

\noindent From the last expression, we can identify the term $\omega= eB/m$ which corresponds to Larmor's frequency. This implies that the (2+1) dimensions Dirac's oscillator represents a charged particle moving in a plate in presence of a perpendicular magnetic field to this particle. Using the result  (\ref{eqE54.1}), we study the lagrangian density of the system 

\begin{eqnarray}\nonumber
\mathcal{L}_{DO}&=& \mathcal{L}_{F} + \mathcal{L}_{I} \\ \nonumber
&=& \bar{\psi} \left(i \gamma^{\mu} \partial_{\mu} - m \right) \psi + \frac{eB}{2} \bar{\psi} \beta \sigma^{\mu \nu} F_{\mu \nu} \psi, \\ \nonumber
&=& \bar{\psi} \left(i \gamma^{\mu} \partial_{\mu} - m \right) \psi + \frac{eB}{2} \bar{\psi} \beta \left( 2 i \alpha_i x_i \right) \psi , \\ \nonumber
&=& \bar{\psi} \left(i \gamma^{\mu} \partial_{\mu} - m \right) \psi + \frac{eB}{2} \bar{\psi} \left( 2 i \alpha_i x_i \right) \psi, \\ \nonumber
&=& \psi^{\dagger} \gamma^0 \left(i \gamma^{0} \partial_{0}+ i \gamma^{j} \partial_{j} - m \right) \psi + \frac{eB}{2} \bar{\psi} \left( 2 i \beta \alpha_i x_i \right) \psi, \\ \nonumber
&=& \psi^{\dagger} \left(i  \partial_{t}+ i \gamma^0 \gamma^{j} \partial_{j} - \gamma^0 m \right) \psi + \frac{eB}{2} \bar{\psi} \left( 2 i \beta \alpha_i x_i \right) \psi, \\ \nonumber
&=& \psi^{\dagger} \left(i  \partial_{t}+ i \alpha_j \partial_{j} - \beta m \right) \psi + \frac{eB}{2} \bar{\psi} \left( 2 i \beta \alpha_i x_i \right) \psi, \\ \nonumber
&=& i \psi^{\dagger} \dot{\psi} + i \psi^{\dagger} \alpha_j \partial_j \psi - \psi^{\dagger} \beta m \psi + \frac{eB}{2} \bar{\psi} \left( 2 i \beta \alpha_i x_i \right) \psi.  \\ \nonumber
&=& i \psi^{\dagger} \dot{\psi} + i \psi^{\dagger} \alpha_1 \partial_x \psi + i \psi^{\dagger} \alpha_2 \partial_y \psi - \psi^{\dagger} \beta m \psi + i  \frac{e B}{m} \psi^{\dagger} \beta \alpha_1 x \psi + i  \frac{eB}{ m} \psi^{\dagger} \beta \alpha_2 y \psi \\
\label{eqE55}
\end{eqnarray}

\noindent From the lagrangian density (\ref{eqE55}) we get the motion equations, using the Euler-Lagrange expresions for the fields $\psi$ and $\psi^{\dagger}$. For the case of the field $\psi$ we have that

\begin{equation}
i \dot{\psi}^{\dagger}= m \psi^{\dagger} \beta - i m \omega \psi ^{\dagger} \beta \alpha_1 x - i m \omega \psi ^{\dagger} \beta \alpha_2 y.
\label{eqE56}
\end{equation}

\noindent whereas for the conjugated field $\psi^{\dagger}$ it is obtained that its motion equation is

\begin{equation}
i \dot{\psi}= \left[\alpha_1 p_1 + \alpha_2 p_2 + m \beta - im \omega \beta \alpha_1 x - im \omega \beta \alpha_2 y  \right] \psi.
\label{eqE57}
\end{equation}

\noindent The expression (\ref{eqE57}) can be rewriten as follows

\begin{equation}
i \dot{\psi}= \left[\alpha_j \left(p_j - i m \omega \beta x_j \right) + \beta m \right] \psi
\label{eqE58}
\end{equation}

\noindent The expressions (\ref{eqE57}) and (\ref{eqE58}) correpond to the Dirac equation in presence of a linear potential in (2+1) dimensions (Dirac's oscillator). On this way, we can conclude that (2+1) dimension Dirac's oscillator represents a particle with 1/2-spin and electrically charged moving inside of a plate interacting with an external and uniform perpendicular magnetic field as we see in the expressions   (\ref{eqBfield2}) and (\ref{eqBfield3}) and also in \cite{4}.

\section{Physical interpretation of (1+1) dimensional Dirac's oscillator}

\noindent To understand the physical meaning of Dirac's oscillator in (1+1) dimensions, lets consider the case of a particle with mass $m$ and electric charge $q$ in presence of a linear electric field $E$ in direction to the $x$-coordinate. The motion equation for this system is \cite{13} \cite{12}
\begin{equation}
q  E = m a,
\label{eqpartlib1}
\end{equation}

\noindent where $a$ is the acceleration of the particle due to its interaction with the electric field. Now, taking into account that the particle moves with uniformly accelerated movement (U.A.M) then its acceleration can be written as follows \cite{26}

\begin{equation}
x= \frac{a t^2}{2} \longrightarrow a= \frac{2x}{t^2}.
\label{eqpartlib2}
\end{equation}

\noindent Replacing (\ref{eqpartlib2}) in (\ref{eqpartlib1}), we have

\begin{equation}
E= \frac{m}{q} \left(\frac{2x}{t^2} \right).
\label{eqpartlib3}
\end{equation}

\noindent Also, taking into account that the velocity of a particle in an accelerated motion can be written as

\begin{equation}
v=at  \longrightarrow t= v/a,
\label{eqpartlib4}
\end{equation}

\noindent where we have taken $v_0=0$. In this way, substituting  (\ref{eqpartlib4}) in (\ref{eqpartlib3}) then

\begin{equation}
E= \frac{2m}{q} \frac{a^2}{v^2} x
\label{eqpartlib5}
\end{equation}

\noindent In order to make the connection between the presented situation in this section and the (1+1) dimension Dirac's oscillator physical interpretation, we write the electric field in this situation explicitly \cite{10} \cite{13} \cite{12}

\begin{equation}
\vec{E}= - \zeta x \hat{x}, 
\label{eqE59}
\end{equation}

\noindent where $\zeta$ is a paremeter which will be calculate later. Whereas the magnetic field is zero in all the space, i.e. 

\begin{equation}
\vec{B}=0.
\label{eqE60}
\end{equation}

\noindent In this case, we consider that the source of the electric field is in the infinite to neglect the edge effects. In the charge's outer reference frame, the 4-electromagnetic potential can be written as follows \cite{11}

\begin{equation}
A^{\mu}= \zeta \left(0, tx \right).
\label{eqE61}
\end{equation}

\noindent In the same way that the (2+1) dimensional case, we shall study the electromagnetic fields produced by this 4-potential

\begin{eqnarray}\nonumber
\vec{E}&=& - \vec{\nabla} \phi - \partial_t \vec{A}, \\ \nonumber
&=& - \vec{\nabla} \left(0 \right) - \partial_t \left(\zeta x t \right), \\ \nonumber
\end{eqnarray}

\begin{equation}
\vec{E}= - \zeta x.
\label{eqE62}
\end{equation}

\begin{eqnarray}\nonumber
\vec{B}&=& \vec{\nabla} \times \vec{A} \\ \nonumber
&=& \left| \begin{array}{ccc}
\hat{x}& \hat{y}& \hat{z}\\
\partial_{x}& \partial_{y}&\partial_{z}\\
0 & 0 &0\\
\end{array}\right|. \\ \nonumber
\end{eqnarray}

\begin{equation}
\vec{B}= 0.
\label{eqE63}
\end{equation}

\noindent We see that (\ref{eqE62}) and (\ref{eqE63}) are equal than (\ref{eqE59}) and (\ref{eqE60}).  The idea is to express these results in a covariant form. For that reason, we take advantage to the fact that electromagnetic fields are gauge invariant and, hence we choose the gauge function

\begin{equation}
\Lambda= - \frac{\zeta}{4} \left(t x^2 + \frac{t^3}{3} \right).
\label{eqE64}
\end{equation}

\noindent To get the new potential starting from the definition 

\begin{equation}\nonumber
A_{\text{lab}}^{\mu \prime}= A_{\text{lab}}^{\mu} - \partial_{\mu} \Lambda.
\end{equation}

\noindent Now we determine the components both temporal and spatial of the electromagnetic field using the expressions (\ref{eqE61}) and (\ref{eqE64})

\begin{eqnarray} \nonumber
A_{\text{lab}}^{0 \prime}&=& A_{\text{lab}}^{0} - \partial_{t} \Lambda , \\ \nonumber
&=& 0 + \frac{\zeta}{4} \partial_t \left(t x^2 + \frac{t^3}{3} \right), \\ \nonumber
A_{\text{lab}}^{0 \prime}&=& \frac{\zeta}{4} \left(x^2 + t^2 \right). \\
\label{eqE65}
\end{eqnarray}

\begin{eqnarray} \nonumber
A_{\text{lab}}^{1 \prime}&=& A_{\text{lab}}^{1} - \partial_{x} \Lambda , \\ \nonumber
&=& \zeta x t + \frac{\zeta}{4} \partial_x \left(t x^2 + \frac{t^3}{3} \right), \\ \nonumber
A_{\text{lab}}^{1 \prime}&=& \frac{\zeta}{2} x t. \\
\label{eqE66}
\end{eqnarray}

\noindent And in this way the 4-electromagnetic potential of the system is 

\begin{equation}
A_{\text{lab}}^{\mu \prime}= \frac{\zeta}{4} \left(t^2+x^2, 2 t x \right)
\label{eqE67}
\end{equation}

\noindent Now, lets to verify that (\ref{eqE67}) generate the same electric and magnetic fields that describes the physical situation of this section 

\begin{eqnarray}\nonumber
\vec{E}^{\prime}&=& - \vec{\nabla} \phi^{\prime} - \partial_t \vec{A}^{\prime}, \\ \nonumber
&=& - \frac{\zeta}{4} \vec{\nabla} \left(x^2 + t^2 \right)- \frac{\zeta}{4} \partial_t \left(2 t x \right),\\ \nonumber
&=& - \frac{\zeta}{4} \left(2x \right) - \frac{\zeta}{4} \left(2x \right) \\ \nonumber
\vec{E}^{\prime}&=& - \zeta x. \\
\label{eqE68}
\end{eqnarray}

\begin{eqnarray} \nonumber
\vec{B}^{\prime}&=& \vec{\nabla} \times \vec{A}^{\prime}, \\ \nonumber
&=& \left| \begin{array}{ccc}
\hat{x}& \hat{y}& \hat{z}\\
\partial_{x}& \partial_{y}&\partial_{z}\\
0 & 0 &0\\
\end{array}\right|. \\ \nonumber
\vec{B}^{\prime}&=& 0. \\
\label{eqE69}
\end{eqnarray}

\noindent We observed that (\ref{eqE68}) and (\ref{eqE69}) are the same results that we obtained using the 4-potential (\ref{eqE61}). Now, we take the definition (\ref{eqAmu}) to rewrite the electromagnetic potential in a Lorentz's covariant form \cite{9}

\begin{equation}\nonumber
A^{\mu \prime}= \frac{\zeta}{4} \left[2 \left(U \cdot x \right) x^{\mu} - x^2 U^{\mu} \right].
\end{equation}

\noindent Using the definitions(\ref{eqE21}) and (\ref{eqE22}), then the 4-electromagnetic potential is rewritten as 

\begin{equation}
A^{\mu \prime}= \frac{\zeta}{4} \left[2t x^{\mu} -\left(t^2 - x^2 \right)U^{\mu} \right].
\label{eqE70}
\end{equation}

\noindent Whereas that the components of this 4-potential are 

\begin{equation}
A^{0 \prime}= \frac{\zeta}{4} \left(x^2 +  t^2 \right); \hspace{2cm} A^{1 \prime}= \frac{1}{2} \zeta x t.
\label{eqE71}
\end{equation}

\noindent In the same way that (2+1) dimensions, we observe that the definitions (\ref{eqE70}) and (\ref{eqE71}) are identical to the expressions (\ref{eqE65}) and (\ref{eqE66}), and hence the definition (\ref{eqAmu}) also reproduces the results of (1+1) dimensional Dirac's oscillator, which it also implies that this expressions describes the same physical situation that we described at the beginning of this section. Taking into account the latter results, then we can use the definition (\ref{FO2}) for the electromagnetic field tensor \cite{9}

\begin{equation}\nonumber
F_{\mu \nu}= \zeta \left(U^{\mu} x^{\nu} - U^{\nu} x^{\mu} \right).
\end{equation}

\noindent whose non-null components are

\begin{equation}
F_{01}= \zeta x; \hspace{1cm} F_{10}= - \zeta x.
\label{eqE72}
\end{equation}

\noindent Now we shall to connect both situations that we have presented in this section. For that purpose, we can compare (\ref{eqpartlib5}) with $F_{10}$ in (\ref{eqE72}) and we can easily conclude that 

\begin{equation}
\zeta = \frac{2 m a^2}{q v^2}.
\label{zeta1}
\end{equation}  

\noindent Following the same structure than (2+1) dimensional interpretation of Dirac's oscillator, we shall study the quantum behaviour of the particle through of the interaction term  (\ref{eqE54.1}) \cite{15}

\begin{equation}\nonumber
\frac{1}{2} m \omega \sigma^{\mu \nu} F_{\mu \nu}= \frac{1}{2}  \zeta \sigma^{\mu \nu} F_{\mu \nu}.
\end{equation}

\noindent From the last expression we can conclude that $m \omega= \zeta$ and thus $m \omega = \frac{2 m a^2}{q v^2}$ or $\omega = \frac{2 a^2}{q v^2} $. Again, we follow the same procedure that the case in (2+1) dimensions and we study the lagrangian density of this system which has the following form 

\begin{equation}
\mathcal{L}_{DO}= i \psi^{\dagger} \dot{\psi}+ i \psi^{\dagger} \alpha_1 \partial_x \psi  +\psi^{\dagger} \beta m \psi  + i \frac{1}{2} \zeta  \psi^{\dagger} \beta  \alpha_1 x  \psi.
\label{eqE73}
\end{equation}

\noindent Now, we use the Euler-Lagrange equations for the fields $\psi$ and $\psi^{\dagger}$. For the case of the field $\psi^{\dagger}$ we have that

\begin{equation}
i \dot{\psi}^{\dagger}= - \psi^{\dagger} \beta m + i m \omega \psi^{\dagger} \beta \alpha_1 x - i \psi^{\dagger} \alpha_1,
\label{eqE74}
\end{equation}

\noindent whereas that for the conjugated field $\psi^{\dagger}$ it is obtained

\begin{equation}
i \dot{\psi}= \left[\alpha_1 \left(p_1 - i m \omega \beta x \right) + \beta m \right] \psi
\label{eqE75}
\end{equation}

\noindent The expression (\ref{eqE75}) corresponds to the (1+1) dimensional Dirac's oscillator. In this way, we can conclude that the (1+1) dimensional Dirac's oscillator represents a particle moving in a straight line in presence of a linear electric field in the x-coordinate, where $\zeta= - \frac{2m}{q} \frac{a^2}{v^2}$ since that it can act in opposite direction to the particle's displacement. However if we want to see more clear the physical meaning of $\zeta$ in this context, we can take  $F_{01}$ in (\ref{eqE72}) and then derivate it respect its spatial coordinate $x$ and we have

\begin{equation}
\zeta= - \frac{dE}{dx},
\label{zeta2}
\end{equation} 

\noindent and in this way we see clearly that the (1+1) dimensional Dirac's oscillator represents physically a charged particle moving into a linear electric field.

\section{\bf{Lagrangian density of the QED+DO in (2+1) dimensions in its symmetries $U(1)$ y $U(1)_R \times U(1)_L$}}

\noindent In this section, we present the lagrangian density of the Dirac's oscillator in (2+1) dimensions. It is showed that this system which is described by this lagrangian density posses local gauge symmetry $U(1)$, but not local chiral symmetry  $U(1)_R\times U(1)_L$. In the latter case, the Dirac's oscillator is massless because the associated interaction term related with the potential breaks explicitly local chiral symmetry because this term mixes the chirality components. The same results can be obtained for the cases of (3+1) and (1+1) dimensions. 

\subsection{\bf{Local gauge symmetry U(1)}}

\noindent It can be proved that interacting Dirac equation given  by (\ref{eqE56}), for the case in which $\psi(x)$ represents a Dirac's field, can be derived using the Euler-Lagrange equation of the antimatter field $\bar{\psi}(x)$, starting from the lagrangian density of QED+OD defined by

\begin{equation}
\mathcal{L}_{QED+DO}= \mathcal{L}_{D}+\mathcal{L}_{I}+\mathcal{L}_{YM} +eB_I \bar{\psi}(x) \gamma^{0}\gamma^{j}r_j\psi(x), \hspace{0.5cm} j=1,2,
\label{DLQEDOD}
\end{equation}

\noindent where $\mathcal{L}_{D}$ is the free lagrangian density, $\mathcal{L}_{I}$ is the lagrangian density of the interaction and $\mathcal{L}_{YM}$ is the Abelian Yang-Mills lagrangian density. The QED+DO lagrangian density, defined by (\ref{DLQEDOD}), is invariant under local gauge transformations of the group $U(1)$ of the matter fields $\psi(x)$, antimatter fields $\bar{\psi}(x)$ and radiation fields $A_\mu(x)$ which have the following structure

\begin{equation}
\psi(x) \rightarrow \psi^{\prime}(x) = \exp\left[-i \theta(x) \right] \psi(x), \hspace{0.3cm} \bar{\psi}(x) \rightarrow \bar{\psi}^{\prime}(x)= \exp\left[i  \theta(x) \right] \bar{\psi}(x), \label{eqtgQEDmax}
\end{equation}
\begin{equation}
 A_\mu(x) \rightarrow A_\mu^{\prime}(x) = A_\mu -\frac{1}{e}\partial_\mu \theta(x),
\label{eqtgQEDrax}
\end{equation}

\noindent here $x=(y,z,t)$ represents a point in the space-time in (2+1) dimensions. In order to prove the invariance of the lagrangian density (\ref{DLQEDOD}), we write firstly the transformed QED+DO lagrangian density as follows 

\begin{equation}
\mathcal{L}_{QED+OD}^\prime= \mathcal{L}_{D}^\prime+\mathcal{L}_{I}^\prime+\mathcal{L}_{YM}^\prime + eB_I \bar{\psi}^\prime(x) \gamma^{0}\gamma^{j}r_j\psi^\prime(x), \hspace{0.5cm} j=1,2,.
\label{DLQEDODT}
\end{equation}

\noindent After to replace the field of matter, antimatter and radiation given by (\ref{eqtgQEDmax}) and (\ref{eqtgQEDrax}) in (\ref{DLQEDODT}), we have

\begin{equation}
\mathcal{L}_{QED+OD}^\prime= \mathcal{L}_{QED} +eB_I \bar{\psi}(x) \gamma^{0}\gamma^{j}r_j\psi(x), \hspace{0.5cm} j=1,2.
\label{QEDInvGL}
\end{equation}

\noindent The result (\ref{QEDInvGL}) shows the invariance of the lagrangian density of QED+DO under local gauge transformations $U(1)$ of its fields and thus this result implies that this physical system presents local gauge symmetry   $U(1)$.

\subsection{\bf{Explict breaking of the local chiral symmetry  $U(1)_R \times U(1)_L$ of the QED+DO without mass term}}

\noindent In this section, we show that lagrangian density of the QED+DO, for the case of massless Dirac field, does not present a local chiral symmetry $U(1)_R \times U(1)_L$ due to the associated interaction term to the linear potential breaks explicitly the local chiral symmetry since that this term mixes the chirality components. 

\vspace{0.5cm}

\noindent The lagrangian density of the QED+DO in the case of a Dirac's field without mass is given by
 
\begin{equation}
{\mathcal{L}_{QED+DO}}_o= \mathcal{L}_o+\mathcal{L}_{I}+\mathcal{L}_{YM} +eB_I \bar{\psi}(x) \gamma^{0}\gamma^{j}r_j\psi(x), \hspace{0.5cm} j=1,2,
\label{DLQED0DSM}
\end{equation}

\noindent where $\mathcal{L}_o$ represents the free lagrangian density of the massless fermionic field, whereas  $\mathcal{L}_{I}$ and $\mathcal{L}_{YM}$ correspond, again, to the interaction lagrangian density and Yang-Mills lagrangian density, respectively. Writing  $\psi(x)=\psi_R(x)+ \psi_L(x)$, then the expression (\ref{DLQED0DSM}) can be written as

\begin{eqnarray}
{\mathcal{L}_{QED+DO}}_o= \bar{\psi}_R(x) i\gamma^{\mu} \partial_{\mu} \psi_R(x) + \bar{\psi}_L(x) i\gamma^{\mu} \partial_{\mu} \psi_L(x)-e \bar{\psi}_R(x) \gamma^{\mu} A_{\mu} \psi_R(x) -e\bar{\psi}_L(x) \gamma^{\mu} A_{\mu} \psi_L(x) \nonumber \\+ eB_I \bar{\psi}_R(x) \gamma^{0}\gamma^{j}r_j\psi_L(x) + eB_I \bar{\psi}_L(x) \gamma^{0}\gamma^{j}r_j\psi_R(x) ,
\label{DLQEDODSM}
\end{eqnarray}

\noindent in the last expression we take into account that the terms $\bar{\psi}_R(x) \gamma^{\mu} \partial_{\mu} \psi_L(x)$,  $\bar{\psi}_L(x) \gamma^{\mu} \partial_{\mu} \psi_R(x)$, $\bar{\psi}_R(x) \gamma^{0}\gamma^{j}\psi_R(x)$ and  $\bar{\psi}_L(x) \gamma^{0}\gamma^{j}\psi_L(x)$ vanish because $P_R P_L  = P_L P_R  =0$. Next, it is proved that the lagrangian density  (\ref{DLQEDODSM}) does not present the local chiral symmetry   $U(1)_R \times U(1)_L$, because it is not invariant under local phase transformations of the chirality projections of the matter, antimatter and radiation fields which have the following form 

\begin{equation}
\psi_L(x) \rightarrow \psi^{\prime}_L(x)=\exp\left[- i \theta_L(x) \right] \psi_L (x);\hspace{1cm} \psi_R(x) \rightarrow \psi^{\prime}_R(x)=\exp\left[- i \theta_R(x)  \right] \psi_R(x),\label{TGLCMx}
\end{equation}
\begin{equation}
\bar{\psi}_L(x) \rightarrow \bar{\psi}^{\prime}_L(x)=\exp\left[i \theta_L(x) \right] \bar{\psi}_L (x); \hspace{1cm} \bar{\psi}_R(x) \rightarrow \bar{\psi}^{\prime}_R(x)=\exp\left[i \theta_R(x)  \right] \bar{\psi}_R(x),\label{TGLCAx}
\end{equation}
\begin{equation}
A_\mu(x) \rightarrow {A_\mu^L}^\prime(x)=A_\mu(x)-\frac{1}{e}\partial_\mu \theta_L(x); \hspace{1cm} A_\mu(x) \rightarrow {A_\mu^R}^\prime(x)=A_\mu(x)-\frac{1}{e}\partial_\mu \theta_R(x),\label{TGLCRx}
\end{equation}

\noindent where $\theta_R(x)$ and $\theta_L(x)$ are different gauge functions associated to the symmetry groups $U(1)_R$ and $U(1)_L$. To prove that this lagrangian does not have invariance, we firstly write a transformed lagrangian density of QED+DO for the massless Dirac's oscillator as follows 

\begin{equation}
{\mathcal{L}^\prime_{QED+DO}}_o= \mathcal{L}_o^\prime+\mathcal{L}_{I}^\prime+\mathcal{L}_{YM}^\prime+ eB_I \bar{\psi}^\prime_R(x) \gamma^{0}\gamma^{j}r_j\psi^\prime_L(x) + eB_I \bar{\psi}^\prime_L(x) \gamma^{0}\gamma^{j}r_j\psi^\prime_R(x).
\label{DLQEDODSMT}
\end{equation}

\noindent After to replace the transformed fields given by ($\ref{TGLCMx}$), ($\ref{TGLCAx}$) and ($\ref{TGLCRx}$) in ($\ref{DLQEDODSMT}$), we have that

\begin{eqnarray}
{\mathcal{L}^\prime_{QED+DO}}_o &=&\mathcal{L}_o+\mathcal{L}_{I}+\mathcal{L}_{YM}+ eB_I \left(\exp \left[ i\theta_R(x)-i\theta_L(x) \right] \bar{\psi}_R(x) \gamma^{0}\gamma^{j}r_j\psi_L(x) \right) \nonumber\\ &+& eB_I \left(\exp \left[ i\theta_L(x)-i\theta_R(x) \right]\bar{\psi}_L(x) \gamma^{0}\gamma^{j}r_j\psi_R(x)\right)\neq{\mathcal{L}_{QED+OD}}_o,
\label{DLQEDIQ}
\end{eqnarray}

\noindent due to $\theta_R (x) \neq \theta_L (x)$, we have that $\theta_R(x) - \theta_L(x)\neq 0$ and $\theta_L(x) - \theta_R(x) \neq 0$. For those reasons, we have shown that massless Dirac's oscillator interacting with an external and uniform magnetic field does not have local chiral symmetry $U(1)_R \times U(1)_L$, because of the associated interaction term to the linear potential breaks explicitly this local chiral symmetry.

\section{Conclusions}

\noindent We have seen that (2+1) dimensional Dirac's oscillator physically corresponds to a particle with 1/2-spin and electrically charged moving inside of a plate interacting with an external and uniform perpendicular magnetic field whereas the (1+1) dimensional Dirac's oscillator, represents an electrically charged particle moving into a linear electric field (respect its space coordinate). This means that when the dimensionality of the Dirac's oscillator changes, the physical system changes completely. This is an unusual result because there are few physical systems which change totally when its spatial dimension changes and for this reason, Dirac's oscillator is more interesting to study either theoretically and experimentally and even from computational physics.

\vspace{0.5cm}

\noindent It is also observed that independently of its dimensionality, (2+1) or (1+1), Dirac's oscillator preserves local gauge symmetry $U(1)$ but the chiral symmetry  $U(1)_R \times U(1)_L$ is not preserved in both cases. This means that despite that (2+1) and (1+1) dimensional Dirac's oscillator represents different physical systems, both have the same symmetries preserved and broken. These results can be explained because the form of the lagrangian (and therefore of the motion equations for the fields $\psi$ and $\psi^{\dagger}$) for (1+1) and (2+1) dimensional Dirac's Oscillator are similar.  

\vspace{0.5cm}

\noindent Finally, we can see that (2+1) and (1+1) dimensional Dirac's oscillator represents electrodynamical systems, and the same happen with (3+1) dimensional case \cite{2} which represents to a particle without charge moving with anomalous magnetic moment inside of a dielectric sphere interacting with an external electric field which depends with its radial distance \cite{2}. Both results are similar because the method to study them is not too different that the authors applied in (3+1) case because their lagrangian densities are very similar mathematically.

\end{document}